\documentclass[twocolumn,prl,amsmath,amssymb,superscriptaddress]{revtex4}
\usepackage{graphicx}
\usepackage{amsmath}
\usepackage{amssymb}
\usepackage{dcolumn}
\usepackage{color}
\usepackage{tikz}
\usepackage{braket}
\usepackage[normalem]{ulem}

\begin{document}

\title{Mean-field scaling of the superfluid to Mott insulator transition in a two-dimensional optical superlattice}

\author{Claire K. Thomas}
\author{Thomas H. Barter}
\author{Tsz-Him Leung}
\author{Masayuki Okano}
\author{Gyu-Boong Jo}
\altaffiliation[Present address: ]{Department of Physics, Hong Kong University of Science and Technology, Clear Water Bay, Kowloon, Hong Kong}
\author{Jennie Guzman}
\altaffiliation[Present address: ]{Sandia National Laboratories, Livermore, CA 94550, USA; and Department of Physics, California State University, Hayward, CA 94542, USA}
\author{Itamar Kimchi}
\altaffiliation[Present address: ]{Department of Physics, Massachusetts Institute of Technology, Cambridge, MA 02139, USA}
\author{Ashvin Vishwanath}
\altaffiliation[Present address: ]{Department of Physics, Harvard University, Cambridge, Massachusetts 02138, USA}
\affiliation{Department of Physics, University of California, Berkeley, CA 94720, USA}
\author{Dan M. Stamper-Kurn}
\affiliation{Department of Physics, University of California, Berkeley, CA 94720, USA}
\affiliation{Materials Sciences Division, Lawrence Berkeley National Laboratory, Berkeley, CA 94720, USA}

\date{\today }

\begin{abstract}
The mean-field treatment of the Bose-Hubbard model predicts properties of lattice-trapped gases to be insensitive to the specific lattice geometry once system energies are scaled by the lattice coordination number $z$.
We test this scaling directly by comparing coherence properties of $^{87}$Rb gases that are driven across the superfluid to Mott insulator transition within optical lattices of either the kagome ($z=4$) or the triangular ($z=6$) geometries.  The coherent fraction measured for atoms in the kagome lattice is lower than for those in a triangular lattice with the same interaction and tunneling energies.  A comparison of measurements from both lattices agrees quantitatively with the scaling prediction.  We also study the response of the gas to a change in lattice geometry, and observe the dynamics as a strongly interacting kagome-lattice gas is suddenly ``hole-doped" by introducing the additional sites of the triangular lattice. 
\end{abstract}


\maketitle

The Bose-Hubbard model describes bosons confined to a lattice, and predicts a low-temperature phase transition between superfluid and Mott insulating states that is driven by on-site interactions \cite{Fisher1989}.
A mean-field treatment of this model neglects non-local correlations and specifies that system properties such as particle number ($n$), superfluid number ($n_\text{sf}$), and entropy ($s$) per lattice site depend on the system's characteristic energies -- the chemical potential  ($\mu$), on-site interaction energy ($U$), and thermal energy ($\tau = k_B T$) -- once they are scaled by $z J$, where $z$ is the lattice coordination number and $J$ is the tunneling energy. 
Aside from the inclusion of $z$, the mean-field theory is insensitive to the lattice geometry.  
Treatments that consider non-local correlations deviate from mean-field predictions, particularly in low-dimensional systems \cite{Capogrosso-Sansone2007,  Sheshadri1995, DosSantos2009, Pai2012, Lin2012, Wei2016, Freericks1996, Li2012}. 

Ultracold Bose gases trapped in optical lattices realize the Bose-Hubbard model \cite{jaks98lattice,grei02mott} and have allowed for experiments that identify the zero-temperature critical point with moderate precision by measuring either the fraction of atoms at zero quasimomentum \cite{Spielman2008,Jimenez-Garcia2010} or the closing of the Higgs-mode energy gap \cite{endr12higgs} in two-dimensional (2D) square lattices. 
The observed critical interaction strengths range between the mean-field prediction and the higher value predicted by more advanced methods \cite{Capogrosso-Sansone2007}.
However, the interpretation of measurements is complicated by the non-zero temperature of the gas, which is difficult to determine and control accurately \cite{trot10temp}, and by its external harmonic confinement, which causes local system properties to vary spatially \cite{Rigol2009}.

Here, we propose and pursue an alternate approach wherein we directly test the mean-field scaling prediction for the Bose-Hubbard model.  Unlike previous tests, ours does not require identifying the precise critical point.  Our method applies regardless of the temperature of the gas, and remains valid even when the exact temperature is not known.  Moreover, our test is valid in the presence of external harmonic confinement.

Our test compares bulk properties of trapped gases that are prepared at the same total particle number $N$ and total entropy $S$, but within optical lattices with two different coordination numbers.  Under the hypothesis that system properties are determined locally, i.e., using both the local density and mean-field approximations, global system properties such as $N$, $S$, and the total superfluid population $N_{\text{sf}}$ are determined by integrating over a three-dimensional trapped sample as
\begin{equation}
F = K \int_{-\infty}^{\tilde{\mu}} d\tilde{\mu}^\prime \, \sqrt{\tilde{\mu} - \tilde{\mu}^\prime} \, f(\tilde{\mu}^\prime, \tilde{U}, \tilde{\tau}),
\end{equation}
where $F \in \{N, N_{\text{sf}}, S\}$,  $f \in \{n, n_{\text{sf}}, s\}$  and the tilde indicates an energy scaled by $z J$.  The effective number of occupied lattice sites is given by  $K = (\pi \alpha/v \bar{\omega}^3) (2 z J/m)^{3/2}$, 
where $\alpha$ is the number of equivalent sites in the unit cell, $v$ is the unit cell volume, $m$ is the atomic mass, and $\bar{\omega}$ is the geometric mean trapping frequency.  The quantity $N/K$ generalizes the ``characteristic density,'' defined in Ref.\ \cite{Rigol2009}, to non-square lattices.

According to the equations above, if the effective site numbers $K$ and the scaled interaction energies $\tilde{U}$ are the same for samples confined to lattices of different $z$ and with the same $N$ and $S$, then  $\tilde{\mu}$, $\tilde{\tau}$, and, consequently, $N_\text{sf}$ will be the same in the two lattices regardless of the specific functional form of $f (\tilde{\mu}^\prime, \tilde{U}, \tilde{\tau})$.  This prediction applies at both zero and non-zero temperature.  In contrast to previous experiments and many numerical treatments, in our test, the coherence properties of the Bose-Hubbbard model for different lattice geometries are compared not just at a critical point, but along a line in the $\tilde{\mu}-\tilde{\tau}$ plane.

We perform this test experimentally by exploring the phase transition between the superfluid and Mott insulating states of $^{87}$Rb atomic gases in the 2D triangular and kagome geometries, which have coordination numbers $z_\mathrm{tri}=6$ and $z_\mathrm{kag} = 4$.  While the measured data for each lattice agree roughly with mean-field predictions, we focus instead on testing the scaling hypothesis by comparing measurements from both lattices over a range of $\tilde{U}$.  Our results agree quantitatively with the scaling prediction.

Finally, we explore the dynamic response of the quantum gas to a change in lattice geometry.  We prepare a gas in the kagome lattice and, then, quench the lattice to the triangular geometry on timescales either fast or slow compared to the characteristic tunneling time, $h/zJ$.  A slow quench increases the coherent fraction, while a rapid quench, in which the gas is suddenly ``hole-doped'' by the addition of vacant lattice sites, results in transient dynamics that damp out and lead to heating. 


We use an optical superlattice created by overlaying two triangular lattices, each formed at the intersection of three beams of light of equal intensity that intersect at equal angles in the horizontal plane, and have in-plane polarization \cite{jo12kag}.  We use light at $532$ nm and $1064$ nm wavelengths, resulting in lattices with intensity minima spaced by $a = 355$ nm and $2a$, respectively. The 532-nm light, blue-detuned from the principal atomic resonances,  attracts atoms to its intensity minima.  The resulting triangular lattice potential has a depth $V_{532}$ that determines $U$ and $J$, where $U$ also depends on the depth $V_{\perp}$ of an additional vertical lattice.

\begin{figure}[t]
\centering
\includegraphics[]{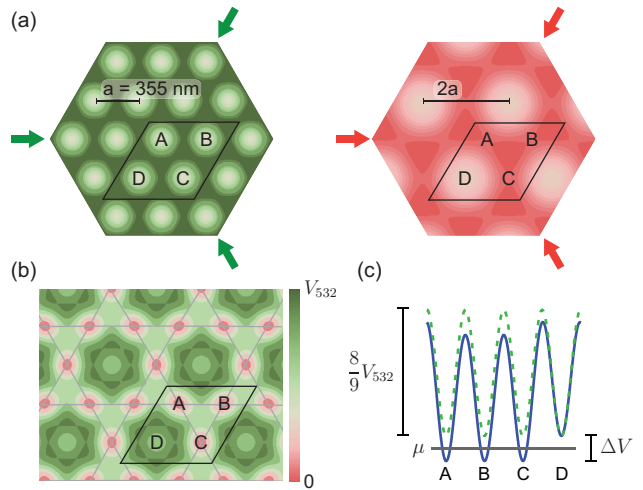}
\caption{Triangular and kagome optical lattices.  (a) Triangular lattices created by light at wavelengths 532 nm (left) and 1064 nm (right).  Intensity (indicated by color saturation) minima are separated by lattice spacings $a = 355$ nm and $2a$, respectively. 
(b) Overlapping the 1064-nm intensity minima with those of the 532-nm lattice (sites D) yields a kagome lattice.
(c)   Potential energy along a path between sites of the unit cell: 532-nm lattice only (green dashed line) and bichromatic lattice (blue line).  Atoms are confined to the kagome geometry when $\Delta V$ exceeds the chemical potential $\mu$ and the tunneling energy $J$.
}
\label{fig:lattice}
\end{figure}

A unit cell of the 1064-nm lattice contains four sites of the 532-nm lattice, labeled A -- D in Fig.\ \ref{fig:lattice}.  At low intensity, the 1064-nm lattice primarily introduces energy offsets $V_{A,B,C,D}$ among the four sites in the unit cell and has little influence on $U$ or $J$ \cite{note:UJconstant}.  When the 1064-nm intensity minima coincide with 532-nm intensity minima (sites D in Fig.\ \ref{fig:lattice}), the superlattice unit cell has three degenerate low-energy sites, and one high-energy site offset by an energy $\Delta V = V_D - V_{A,B,C} = 8/9 \times V_{1064}$.  When $\Delta V$ exceeds the relevant energies ($\mu$ and $J$) of low-temperature atoms in the ground band of the lattice, the atoms become restricted to the kagome lattice.  Our kagome-lattice data are taken with $\Delta V/h =  13 \, \mbox{kHz}$, which satisfies the stated criteria as the chemical potential ranges between $\mu/h =$ 1.5 and 2.9 kHz.  The relative position of the commensurate lattices is measured interferometrically and stabilized actively.

\begin{figure}[t]
\centering
\includegraphics[]{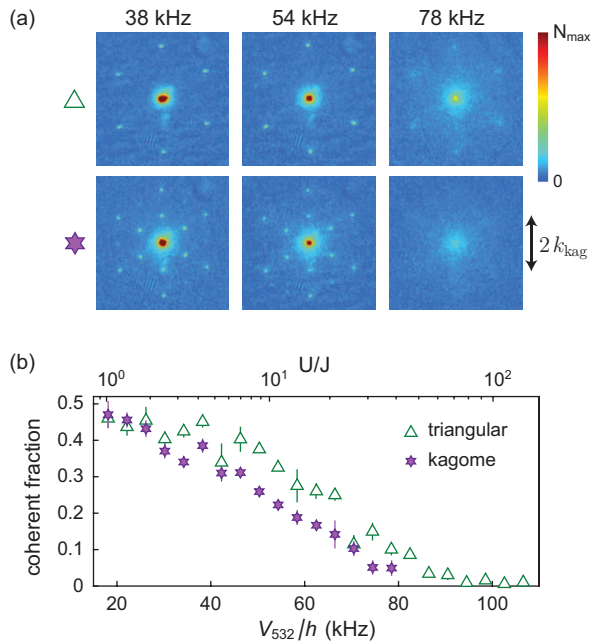}
\caption{Comparing the superfluid to Mott insulator transition in lattices with different coordination numbers.
(a) Momentum-space images of atoms released from the triangular (upper row) and kagome (lower row) lattices for $V_{532}/h = 38, 54 \text{ and } 78$ kHz.  
Diffraction of the superfluid produces sharp peaks at reciprocal lattice vectors; those from kagome lattice show additional peaks at wavenumber $k_{\mathrm{kag}} = 2 \pi \sqrt{3}/(1064 \mbox{nm})$.  On-site interactions in deeper lattice drive a phase transition to the Mott insulating state, as indicated by loss of coherence.
(b) We measure coherent fraction by summing the populations at all coherent diffraction peaks and dividing by the total atom number.  Data are shown as a function of $V_{532}$ (lower axis) and of $U/J$ (upper axis). Each point represents 3-5 experimental iterations and the estimated standard error in the mean for each point is shown.  See Ref.\ \cite{beck10tri} for measurements in triangular lattices at higher filling.}
\label{fig:sfMI_UJ}
\end{figure}

For our experiments, we prepare nearly pure $^{87}$Rb Bose-Einstein condensates of between $0.5$ and $3 \times 10^{5}$ atoms in the $\ket{F=1,m_F=-1}$ hyperfine state in a red-detuned crossed optical dipole trap, characterized by trap frequencies of $(\omega_x, \omega_y, \omega_z)=2\pi \times (34, 64, 49)$ Hz.  We then impose a one-dimensional lattice with potential depth $V_{\perp}/h= 41$  kHz formed by a retro-reflected 1064-nm-wavelength beam propagating vertically.  The gas becomes divided among $\simeq 17$ decoupled planes (with a single-atom tunneling rate of 5 Hz).

Releasing this gas from its confinement, allowing time of flight, and imaging in the horizontal plane reveals a coherent fraction of around 0.5, far lower than observed before applying the vertical lattice \cite{note:depletion}.  We ascribe this reduction to the effect of elastic collisions between vertically expanding portions of the gas that transfer momentum incoherently into the horizontal direction, leading to an underestimate of the coherent fraction in the lattice.

With the vertical lattice in place, we load the atoms into the 2D superlattice with a simultaneous increase of the superlattice beam intensities  \cite{note:loading}. 
After allowing the gas to evolve at the final lattice depths for 30 ms, we release the atoms into a loose horizontally confining magnetic potential in which they undergo a quarter-cycle of motion before we probe them by absorption imaging in the horizontal plane for a momentum-space characterization of the gas \cite{MomFocus}. The vertical lattice is ramped off 150 $\mu$s before the superlattice and optical traps to reduce the collisional transfer of vertical to transverse momentum.

As shown in typical momentum-space measurements [Fig.\ \ref{fig:sfMI_UJ}(a)], as either lattice is deepened, coherent diffraction peaks give way to a diffuse momentum distribution that represents both the incoherent portion of the gas and the effects of elastic scattering during expansion.  We quantify the coherent fraction [Fig.\ \ref{fig:sfMI_UJ}(b)] by using bimodal 2D fits to count the number of atoms in each sharp diffraction peak above the diffuse background and dividing by the total number of atoms in an image.

The momentum-space images and resulting coherent fraction measurements show the influence of lattice geometry on the properties of ultracold bosons trapped within the lattice.  At all lattice depths, the superfluid is less robust in the kagome lattice than in the triangular lattice, as expected owing to the lower coordination number.


Given our experimental parameters, we expect an $n = 2$ Mott insulator to form at the center of our gas.  We observe the coherent fraction becoming negligible (below a few percent and consistent with zero) near $U/J = 60$ for the triangular lattice and $U/J = 40$ for the kagome lattice.  Both values are consistent with the mean-field prediction that the critical point for forming the $n=2$ Mott insulator lies at $\tilde{U}= 9.9$ at low entropy.  However, the trap inhomogeneity and, in particular, our uncertainty about the entropy of the gas, which we estimate as $S/N < 0.3 \, k_B$ based on comparisons to numerical calculations (see below), preclude a more precise comparison to the predicted critical values. 
 
\begin{figure}[t]
\centering
\includegraphics[]{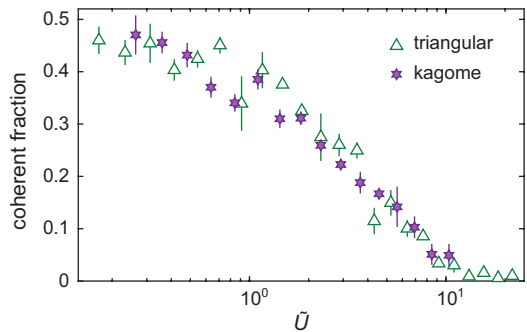}
\caption{A test of the scaling hypothesis in which the coherent fractions measured in the triangular and kagome lattices [Fig.\ \ref{fig:sfMI_UJ}(b)]
are plotted against the scaled interaction energy $\tilde{U} = U/zJ$.  The overlap of the two data sets at all $\tilde{U}$ indicates agreement with the scaling prediction.}
\label{fig:UJz}
\end{figure}


However, rather than focusing on an imprecisely measured location of a critical point, our main objective is to utilize our measurements for a precise test of the scaling hypothesis. We emphasize again that this test remains valid even in the absence of a precisely determined entropy for the gas under study. For this test, we compare the coherent fraction of atoms measured from either the triangular or kagome lattices.  We observe that scaling the experimental $U/J$ by $z^{-1}$ leads to very good overlap between the two datasets, within experimental error at all values of $\tilde{U}$  (Fig.\ \ref{fig:UJz}). 


More quantitatively, we utilize the entire dataset at all values of $\tilde{U}$  to determine the factor $\zeta$ by which the $U/J$ axis of the kagome-lattice dataset should be scaled to best fit the triangular-lattice dataset.  This is done by applying simultaneous spline fits to the two datasets, and then determining the value of $\zeta$ by an error-weighted least-squares measure.  We obtain $\zeta = 1.6(1)$, where scaling predicts $\zeta = z_\text{tri}/z_\text{kag}= 1.5$.

Therefore, within its 6\% estimated error, our measurement supports the scaling  predicted by mean-field theory.  However, our data do not necessarily disagree with beyond-mean-field theories.  Critical values of $U/J$ for the formation of both the $n=1$ and $n=2$ Mott insulators in a zero-temperature homogeneous system have been calculated using a high-order perturbation method.  These calculated critical values in the 2D triangular and kagome lattices have the ratio 1.65, in disagreement with the scaling prediction \cite{lin12kagmott}.  Yet, we cannot compare these calculations directly to our experimental findings because they do not account for inhomogeneity or non-zero temperature.  Future comparisons of beyond-mean-field methods to our test are warranted.

We note three imperfections in our approach.  First, for simplicity and for technical reasons,  experiments were performed with a constant trap frequency $\bar{\omega}$.  As a result, the effective site number $K$ in the two lattices differs by the ratio $K_{\text{tri}}/K_{\text{kag}} \simeq 1.4$.  Therefore, the triangular-lattice experiments were performed with a scaled chemical potential $\tilde{\mu}$ and temperature $\tilde{\tau}$ that were both lower than in the kagome-lattice experiments.  In the mean-field picture, data for different $K$ are equivalent to those from experiments performed in the same lattice geometry, but with $N$ and $S$ both scaled by $K^{-1}$, i.e.\ at the same total entropy per particle.  We performed numerical calculations based on non-zero temperature mean-field theory and the local density approximation, and found that scaling both $N$ and $S$ by this amount produced only few-percent changes in the superfluid fraction for gases with small $S/N$ as are used in this experiment.

Second, as discussed above, the coherent fraction determined from our images is an underestimate of the superfluid fraction of the gas.
Our test of the mean-field scaling hypothesis is predicated on the assumption that the systematic underestimation of the coherent fraction is identical for diffraction out of the two different lattices.

Third, the scaling hypothesis applies to a gas at thermal equilibrium.  In our work, we confirmed that the measured coherent fractions were unchanged (at the few percent level) by varying (by factors of two) the times over which the lattice depths were increased and held constant.  We also performed ``round-trip'' measurements, in which, after ramping the lattices on, we ramped off either the horizontal lattice or both the horizontal and vertical lattices and then measured the coherent fraction. In either case, the coherent fraction returned to at least 90\% of its value before the lattices were ramped on. While, this observation constrains the amount by which the gas was heated by application of either the triangular or the kagome lattice, we cannot confirm that the gases studied are at thermal equilibrium.

\begin{figure}[t]
\centering
\includegraphics[]{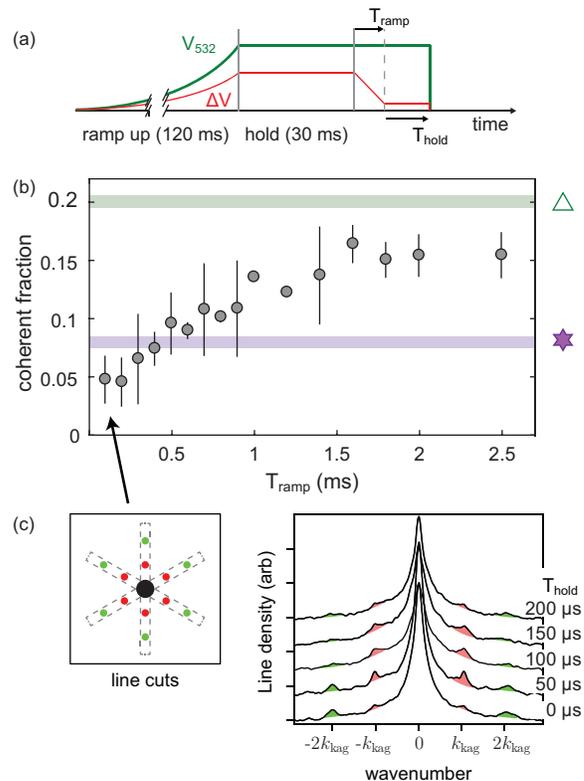}
\caption{Response of a degenerate Bose gas to a structural lattice change. (a) The experimental sequence.  (b) Coherent fraction as a function of ramp time after the gas equilibrates in the final lattice.  Lines show the coherent fraction observed for samples loaded directly into and held within a constant-strength triangular or kagome lattice; line thickness shows the estimated standard error.  (c) Time-resolved response following fastest ramp.  Three one-dimensional line cuts through momentum-space images are summed.  The coherent diffraction at wavenumber $\pm k_{\mathrm{kag}}$ (shaded red to guide the eye) is transiently enhanced as superfluid flows into once-vacant D sites in the unit cell.  First-order diffraction of equilibrated superfluid from triangular lattice occurs at $\pm 2 k_\mathrm{kag}$ (shaded green). 
}
\label{fig:RampTime}
\end{figure}

We also study the evolution of a gas in response to changes in the structure of the optical lattice while $U$ and $J$ remain constant \cite{note:UJconstant}.  We start with a strongly interacting gas with small coherent fraction in the kagome lattice, with $V_{532}/h = 55 \text{ kHz}$  and $\Delta V/h = 15$  kHz, so that $J/h \sim 106 \text{ Hz}$ and $U/h \sim 1.2\, \text{kHz}$.  Note that the entropy per particle of this sample is higher, and thus the coherent fraction lower, than those studied above.  Next, we convert the lattice to the triangular geometry by reducing $\Delta V/h$ linearly in time to a minimal value of $0.5 \text{  kHz}$ in a ramp time $T_\text{ramp}$.  We allow the atoms to evolve for a time $T_\text{hold}$, so that $T_\text{ramp}+T_\text{hold} = 30 \, \mbox{ms}$ is constant, before probing the gas.  

Introducing the additional lattice sites of the triangular lattice in a time that is long compared to the characteristic timescale $h/6 J \sim 1.6$ ms increases the coherent fraction to a near-constant final value (Fig.\ \ref{fig:RampTime}).  That final value is somewhat lower than that observed for a gas loaded directly into a triangular lattice of the same depth and held for an equal total hold time.

More rapid ramps result in a lower coherent fraction.  Through time-resolved measurements [Fig.\ \ref{fig:RampTime}(c)] , we observe that a sudden quench initiates transient dynamics in the strongly interacting superfluid.  These dynamics are evidenced by the redistribution of population among the coherent diffraction peaks that persist for around 200 $\mu$s before the overall coherent fraction decays.




In this work, we made use of an optical superlattice with tunable geometry to study the phase diagram and dynamics of interacting Bose gases in optical lattices.  By comparing the coherence properties of gases in triangular and kagome lattices, we tested whether the local system properties in the different lattice geometries were related simply by a scaling transformation that accounts only for the lattice coordination number.  Our measurements agree precisely with the scaling prediction, placing bounds on the possible influence that band structure effects, such as localization in the low-lying flat band of the kagome lattice, may have on interaction-driven localization in the Mott insulating state.  We also introduce the triangular superlattice as a tool to initiate non-equilibrium dynamics to be studied in future work.

We thank Ehud Altman for helpful discussions.
C.\ K.\ Thomas acknowledges support by the Department of Energy Office of Science Graduate Fellowship Program (DOE SCGF). This work was supported by the NSF and by the AFOSR through the MURI program.

\bibliographystyle{apsrev}
\bibliography{sfMI,MItheorybib,notes}

\begin{thebibliography}{27}
\bibitem[{not({\natexlab{a}})}]{note:UJconstant}
\bibinfo{note}{A steepest-descent minimization algorithm was used to generate
  maximally localized generalized Wannier states \cite{PhysRevA.87.043613}, and
  to determine that that $U$ and $J$ differ by only a few percent in triangular
  and kagome lattices of equal $V_{532}$ depth.}
\bibitem[{not({\natexlab{b}})}]{note:depletion}
\bibinfo{note}{Quantum depletion may account only for a small amount of this
  reduction \cite{mcka15metastable,kato08sharp}}.
\bibitem[{not({\natexlab{c}})}]{note:loading}
\bibinfo{note}{The 532-nm lattice is turned on at $V_{532}/h = 1.7$ kHz to
  initiate stabilization. Its depth is then increased exponentially in time
  (one $e$-fold per 35 ms) until the final depth is reached. For kagome-lattice
  experiments, the 1064-nm lattice is turned on at low intensity and ramped up
  exponentially at a rate adjusted so the two lattices reach their final depths
  simultaneously. $V_{532}$ ($V_{1064}$) was calibrated using
  lattice-modulation spectroscopy in the superfluid regime \cite{ModSpec} and
  was stable within 1\% (6\%) rms over several hours of data taking. Relative
  positions of the lattices were stable within 18 nm rms in each horizontal
  direction.}
\end{thebibliography}

\end{document}